\documentclass[nofootinbib,amsmath,twocolumn,notitlepage,preprintnumbers]{revtex4-1}
\usepackage{multirow}
\usepackage{amssymb, esvect, amsmath, graphicx, latexsym, amsthm, slashed, eso-pic}
\usepackage{xcolor}
\usepackage{hyperref}

    \newcommand{\cH}{{\cal H}}  \newcommand{\cL}{{\cal L}}       
  
\newcommand{\ie}{{\it i.e.}}

\newcommand{\beq}{\begin{equation}} \newcommand{\eeq}{\end{equation}}
\newcommand{\bea}{\begin{eqnarray}} \newcommand{\eea}{\end{eqnarray}}

\def\lsim{\mathrel{\raise.3ex\hbox{$<$\kern-.75em\lower1ex\hbox{$\sim$}}}}
\def\gsim{\mathrel{\raise.3ex\hbox{$>$\kern-.75em\lower1ex\hbox{$\sim$}}}}

\newcommand{\Eq}[1]{Eq.~(\ref{#1})}

\newcommand{\be}{\begin{eqnarray}}
\newcommand{\ee}{\end{eqnarray}}

\newcommand{\benum}{\begin{enumerate}}
\newcommand{\eenum}{\end{enumerate}}
\newcommand{\bi}{\begin{itemize}}
\newcommand{\ei}{\end{itemize}}

\begin{document}

\preprint{FERMILAB-PUB-18-309-A}

\title{WIMPflation}
\author{Dan Hooper$^{a,b,c}$}
\thanks{ORCID: http://orcid.org/0000-0001-8837-4127}
\author{Gordan Krnjaic$^a$}
\thanks{ORCID: http://orcid.org/0000-0001-7420-9577}
\author{Andrew J.~Long$^b$}
\thanks{ORCID: http://orcid.org/0000-0003-0985-5809}
\author{Samuel D.~McDermott$^a$}
\thanks{ORCID: http://orcid.org/0000-0001-5513-1938}

\affiliation{$^a$Fermi National Accelerator Laboratory, Theoretical Astrophysics Group, Batavia, IL, USA}
\affiliation{$^b$University of Chicago, Kavli Institute for Cosmological Physics, Chicago IL, USA}
\affiliation{$^c$University of Chicago, Department of Astronomy and Astrophysics, Chicago IL, USA}

\date{\today}

\begin{abstract}
We propose a class of models in which a stable inflaton is produced as a thermal relic in the early universe and constitutes the dark matter. We show that inflaton annihilations can efficiently reheat the universe, and identify several examples of inflationary potentials that can accommodate all cosmic microwave background  observables and in which the inflaton dark matter candidate has a weak scale mass. As a simple example, we consider annihilations that take place through a Higgs portal interaction, leading to encouraging prospects for future direct detection experiments.

\end{abstract}

\maketitle




\section{Introduction}

Two of the most pressing problems in cosmology concern the unknown physics of dark matter and inflation. Although the existence of dark matter is strongly supported by a variety of observations, the particle identity of this substance remains entirely unknown. Similarly, whereas cosmological inflation is motivated by the flatness and horizon problems~\cite{Guth:1980zm}, and it is supported by the adiabatic and approximately scale-invariant perturbations observed in the cosmic microwave background (CMB)~\cite{Spergel:2006hy,Tegmark:2004qd,Boyle:2005ug,Ade:2015lrj,Ade:2015xua}, we know little about this period of our cosmic history. In this letter, we consider the possibility that these two seemingly unrelated phenomena are in fact intimately connected. More specifically, we explore a broad class of models in which the field responsible for inflation (\ie~the inflaton) is also a stable particle whose population freezes out of thermal equilibrium in the early universe to constitute the dark matter. We refer to this scenario as WIMPflation.  

The possibility that a single particle could play the dual roles of inflaton and dark matter was mentioned in Refs.~\cite{Kofman:1994rk,Kofman:1997yn}, and studied with more detail in Refs.~\cite{Liddle:2006qz,Cardenas:2007xh,Panotopoulos:2007ri,Liddle:2008bm,Bose:2009kc,Lerner:2009xg,Okada:2010jd,DeSantiago:2011qb,Lerner:2011ge,delaMacorra:2012sb,Khoze:2013uia,Kahlhoefer:2015jma,Bastero-Gil:2015lga,Tenkanen:2016twd,Choubey:2017hsq,Heurtier:2017nwl}. In particular, McDonald and collaborators~\cite{Lerner:2009xg,Lerner:2011ge,Kahlhoefer:2015jma} considered a specific and phenomenologically viable model in which the dark matter candidate is also the inflaton, possessing a non-minimal coupling to gravity. In this letter, our goal is to develop a more general outlook on WIMPflation by identifying the features that are generically required of a model that can simultaneously satisfy all existing constraints pertaining to both inflation and dark matter.  We present a wide range of inflationary scenarios in which the inflaton can play the role of a thermal relic that also serves as a viable dark matter candidate.

In order for the inflaton to be a thermal dark matter candidate, it must be stable and it must freeze out of equilibrium in the early universe to yield an acceptable relic abundance. This requires the dark matter to possess interactions that allow it to annihilate with a cross section of approximately $\sigma v \approx 2 \times 10^{-26} \, {\rm cm}^3 / {\rm s}$ and to have a mass between approximately 10 MeV and 100 TeV. Annihilations through these same interactions must take place at the end of inflation and be efficient enough to reheat the universe to a high temperature. This produces a thermal bath of both Standard Model particles and inflatons. The weak scale inflatons then proceed to freeze out of equilibrium in the standard way, resulting in the measured abundance of cold, collisionless dark matter.




\section{Inflationary Dynamics}
\label{dynamics}

We consider a class of scenarios in which the inflaton $\phi$ is a real scalar singlet whose interactions respect a $Z_2$ symmetry, ensuring its stability in the vacuum.  The dynamics of $\phi$ are described by the following Lagrangian:
\be \label{L0}
{\cal L} = \frac{1}{2} g^{\mu \nu} \partial_\mu \phi \, \partial_\nu \phi -V(\phi) + {\cal L}_{\rm int} \, ,
\ee
where $V(\phi)$ is the inflaton potential and ${\cal L}_{\rm int}$ describes the interactions that enable inflaton annihilation to both reheat the universe and later to result in an acceptable thermal relic abundance. 

For the inflaton to be a viable thermal relic, its mass must lie in the range $10 \, {\rm MeV} \lesssim m_\phi \lesssim 100 \, {\rm TeV}$~\cite{Boehm:2013jpa,Griest:1989wd}.  Thus we adopt an inflaton potential that includes one term to generate the inflaton mass plus a second term to drive inflation (without contributing to $m_\phi$). Potentials that satisfy these criteria can be written as 
\begin{eqnarray}
V(\phi) =  \frac{1}{2} \, m^2_{\phi} \phi^2 + \lambda \, \phi^4_0 \, f(\phi/\phi_0) \, , 
\label{Eqpot}
\end{eqnarray}
where $\phi_0$ is a dimensionful constant, the function $f(\phi/\phi_0)$ has a vanishing second derivative at the origin, and thus, $m_{\phi}$ is the inflaton's mass in the vacuum. We consider the following functional forms for the term in the potential that drives inflationary dynamics:
\beq
f(x = \phi/\phi_0) =  \left\{ \begin{array}{l} x^4 \\
\arctan^4 x \\
\tanh^4 x \\
\, \! [1- \exp(-x^2) ]^2  \\
(1-\cos x)^2  \\
x^4/(1+x^2)^2
\end{array} \right. \, . 
\label{sixmodels}
\eeq
The first of these is the $\phi^4$ potential, which is the simplest example that meets the requirements listed above. As is well known, however, $\phi^4$ inflation predicts a large tensor-to-scalar ratio, and is ruled out by modern CMB observations~\cite{Ade:2015lrj,Ade:2015xua}. We include it here for completeness and for an illuminating contrast with the other potentials under consideration.

Several of the examples described in Eq.~(\ref{sixmodels}) are similar to those found in well-known models~\cite{Martin:2013tda}. The function $\arctan(\phi/\phi_0)$ was originally used in Ref.~\cite{Wang:1997cw} as a toy example of a potential that leads to a rapidly varying equation of state parameter. The potentials $\tanh^4(\phi/\phi_0)$ and $[1- \exp(-(\phi/\phi_0)^2)]^2$ are variations of the $T$-model and $E$-model realizations of the $\alpha$-attractor scenario \cite{Kallosh:2015lwa,Kallosh:2013hoa,Kallosh:2013daa,Kallosh:2013yoa}. The potential $[1-\cos(\phi/\phi_0)]^2$ is similar to those found in natural inflation~\cite{Freese:1990rb, Adams:1992bn} but dominated by a higher harmonic~\cite{Kappl:2015esy,Poulin:2018dzj}. Finally, the potential $\phi^4/(1+(\phi/\phi_0)^2)^2$ arises from the $\phi^4$ case in the presence of a non-minimal coupling to gravity of the form $\phi^2 R$~\cite{Bezrukov:2007ep, Lerner:2009xg, Lerner:2011ge}.

As a paradigmatic example for the interaction term in \Eq{L0}, we will consider a Higgs portal operator \cite{McDonald:1993ex,Burgess:2000yq}:
\begin{equation}
 {\cal L}_{\rm int} = -\frac{\kappa}{2} {\mathcal H}^{\dagger}{\mathcal H} \phi^2,
 \label{Lint}
\end{equation} 
where $\kappa$ is a dimensionless mixed quartic coupling and ${\mathcal H}$ is the Standard Model Higgs doublet. In the presence of a background inflaton field, $\phi \neq 0$, the Higgs acquires an effective squared mass given by $m^2_h = -\mu^2_H + \kappa \phi^2/2$ where $\mu_H \sim 10^2 \, {\rm GeV}$ sets the mass of the Standard Model Higgs boson at low temperatures. Over the range of models we will consider here, $m_h$ is always much larger than the rate of Hubble expansion during inflation, and thus quantum fluctuations of the Higgs field are negligible, allowing the dynamics of inflation to be entirely dominated by the second term in Eq.~(\ref{Eqpot}). We also note that ${\mathcal H}=0$ is always a solution to the classical equations of motion, and thus our potential reduces to the case of single-field slow-roll for the purposes of inflationary dynamics.

We emphasize that the second term in Eq.~(\ref{Eqpot}) is entirely responsible for the phenomenology of inflation, whereas Eq.~(\ref{Lint}) and the first term in Eq.~(\ref{Eqpot}) set the inflaton's annihilation cross section and its mass. Therefore, the dark matter and inflationary dynamics are largely modular and independent of one another within this class of models.

If the field $\phi$ is displaced from the minimum of its potential it can drive cosmological inflation and induce the perturbations that we eventually observe in the CMB.  
For each of the functions listed in \Eq{sixmodels}, we calculate the slow-roll parameters, which are defined as follows~\cite{Baumann:2009ds}:
\begin{eqnarray}\label{slowroll}
\epsilon \equiv \frac{1}{2}m^2_{\rm Pl} \bigg(\frac{V'}{V}\bigg)^2, \,\,\,\,\,\,\,\,\,\, \eta \equiv m^2_{\rm Pl} \bigg(\frac{V''}{V}\bigg) \, , 
\end{eqnarray}
where $m_{\rm Pl} \simeq 2.4 \times 10^{18}$ GeV is the reduced Planck mass. Slow-roll inflation occurs when the inflaton field is nearly homogeneous in a Hubble patch and $\epsilon, \eta \ll 1$.  
We define the end of inflation, $t_{\rm end}$, as the time when $\epsilon = 1$.  
CMB observables probe the inflaton potential at $\phi_{\rm CMB}$, which is the value of the inflaton field $N_{\rm CMB} \approx 50$--60 $e$-foldings before the end of inflation:
\begin{equation}
N_{\rm CMB} = \int\limits^{t_{\rm end}}_{t_{\rm CMB}} \! \! H dt = \int\limits^{\phi_{\rm end}}_{\phi_{\rm CMB}} \frac{H d\phi}{\dot{\phi}} = \int\limits^{\phi_{\rm CMB}}_{\phi_{\rm end}} \frac{V}{V'} \frac{d\phi}{m^2_{\rm Pl}} \, . 
\end{equation}
In the last step we have used the slow-roll approximation, $\dot{\phi} = -V'/3H$.


\begin{figure}
\includegraphics[width=3.4in,angle=0]{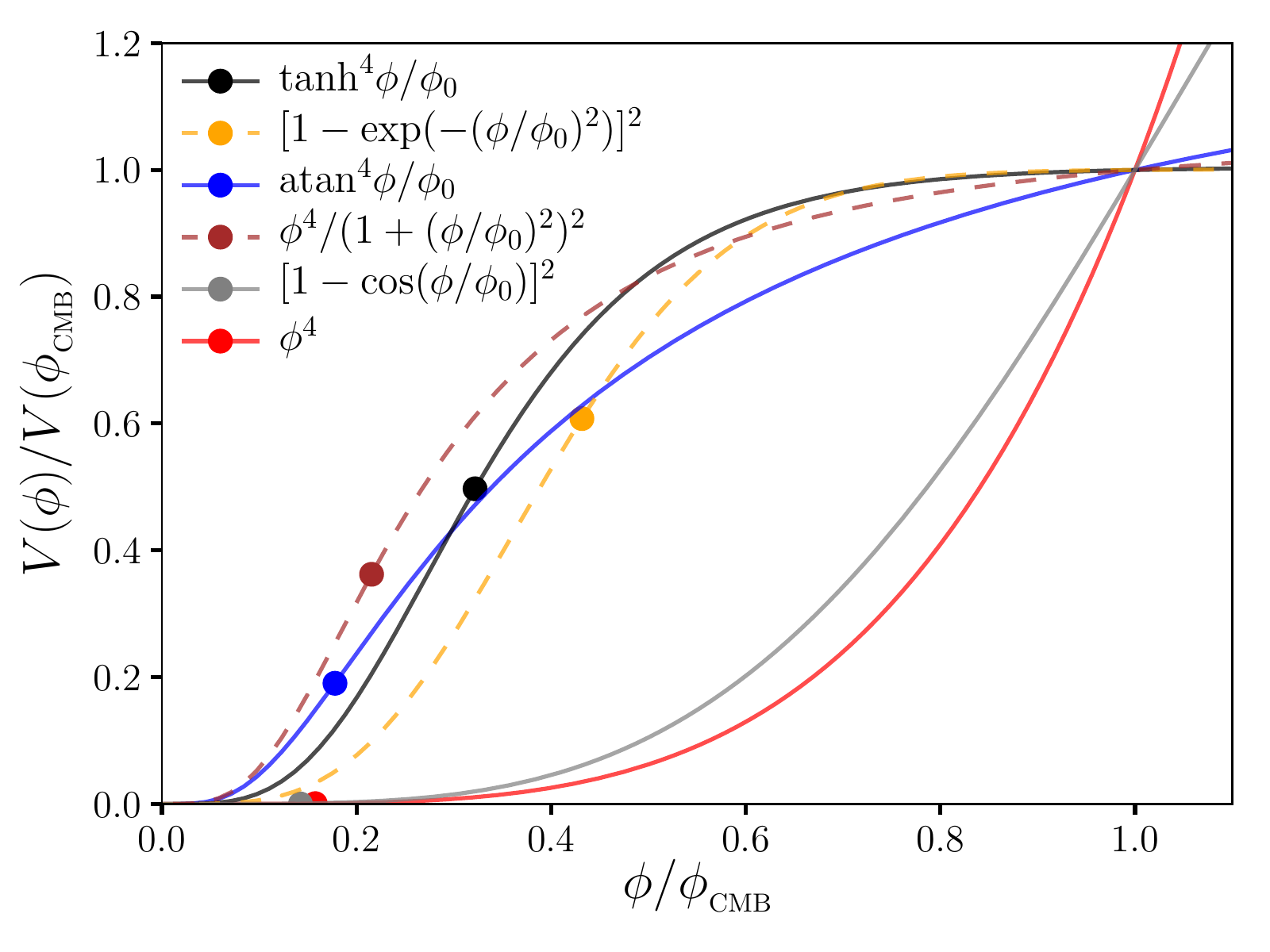}
\caption{The inflaton potentials considered in this study are shown here normalized to the value that they take 60 $e$-foldings before the end of inflation. 
The colored dots denote the point at which inflation ends in each model.}
\label{potential}
\end{figure}


\begin{figure}
\includegraphics[width=3.4in,angle=0]{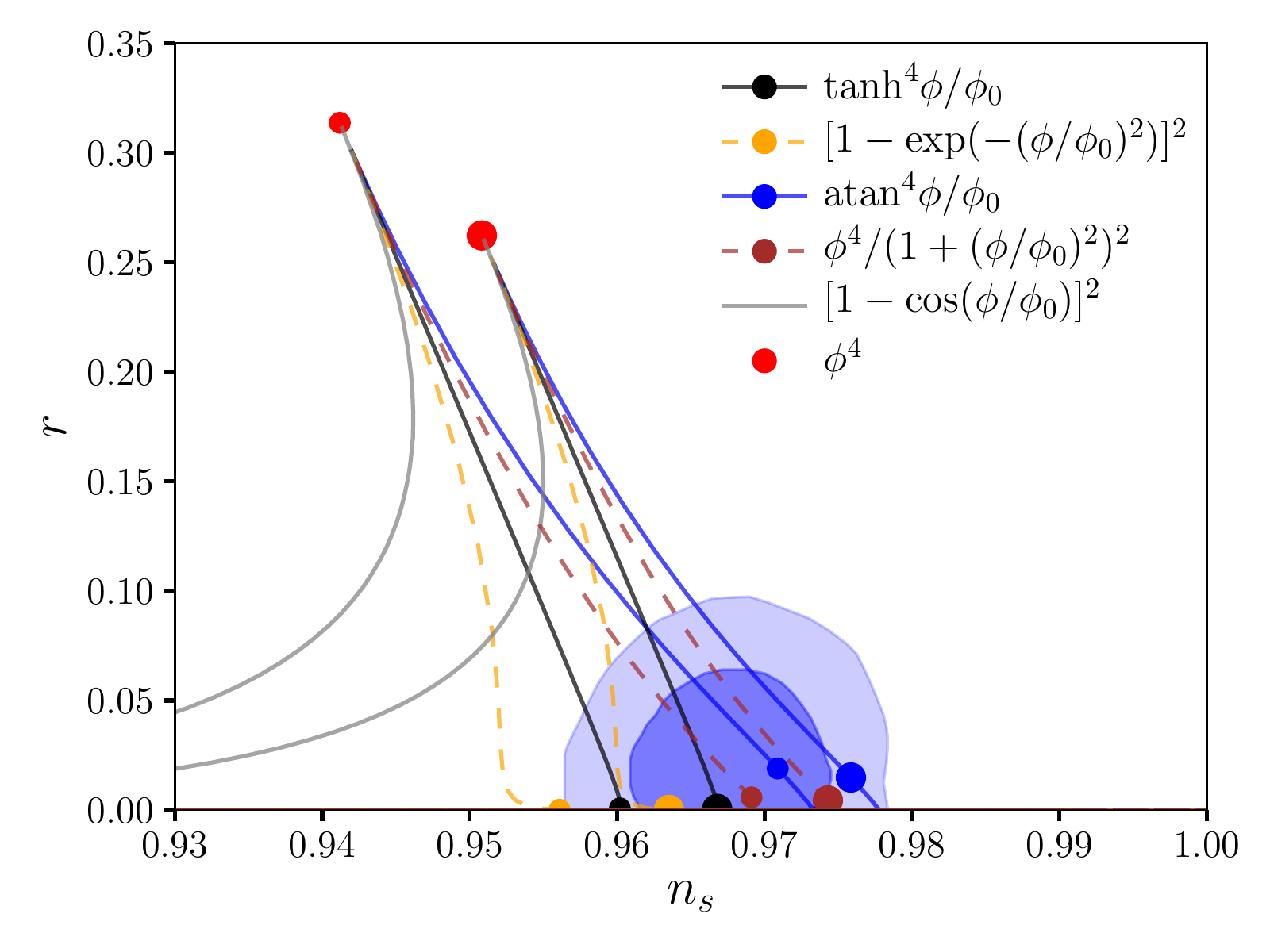}
\caption{The tensor-to-scalar ratio ($r$) and the tilt of the scalar power spectrum ($n_s$) predicted for each of the inflationary potentials considered in this study. Along each curve, the value of $\phi_0$ varies, and the left and right curves correspond to results found for $N_{\rm CMB}=50$ and 60, respectively. The colored dots denote the point at which $\phi_0=m_{\rm Pl}$ (no grey point is shown, as $\phi_0 > m_{\rm Pl}$ over the entire range shown for this model). Planck measurements (68 and 95\% confidence contours shown in dark and light blue, respectively)~\cite{Ade:2015lrj} 
disfavor the $\phi^4$ and $[1-\cos (\phi/\phi_0)]^2$ models.}
\label{planck}
\end{figure}

In Fig.~\ref{potential}, we plot representative examples of the six WIMPflaton potentials being considered in this study, normalized to their value at $\phi_{\rm CMB}$ and for the case of $N_{\rm CMB}=60$. For each of these curves, we have adopted a value of $\phi_0=m_{\rm Pl}$, except for in the $[1-\cos (\phi/\phi_0)]^2$ case, for which we show results for $\phi_0=10 m_{\rm Pl}$ (the slow-roll conditions can only be satisfied in this case if $\phi_0 \gg m_{\rm Pl}$). The colored dots along each curve denote the point at which $\phi=\phi_{\rm end}$.

Next, we calculate the values of the tensor-to-scalar ratio, $r = 16 \epsilon$, and the tilt of the scalar power spectrum, $n_s=1+2\eta-6\epsilon$, predicted in this class of models (where $\epsilon$ and $\eta$ are evaluated at $\phi_{\rm CMB}$), and compare these quantities to the constraints imposed by the Planck Collaboration~\cite{Ade:2015lrj}. As can be seen in Fig.~\ref{planck}, the predictions of the $\phi^4$ and $[1-\cos (\phi/\phi_0)]^2$ models are in considerable tension with the data. The other four models, however, yield predictions that can easily accommodate the measurements from Planck.  In the large $\phi_0$ limit, each of these potentials reproduces the $\phi^4$ prediction.

For the class of potentials in Eq.~(\ref{sixmodels}), we roughly have $V(\phi_{\rm CMB}) \sim \lambda m^4_{\rm Pl}$ during slow roll, so the
amplitude of the scalar power spectrum satisfies 
\be\label{eq:As}
A_s = \frac{V(\phi_{\rm CMB})}{24 \pi^2 m^4_{\rm Pl} \epsilon} \sim 2 \times 10^{-9} \left( \frac{\lambda}{5 \times 10^{-9}} \right) \left( \frac{10^{-2}}{\epsilon} \right).~~
\ee
Thus $\lambda$ must be very small in order to accommodate Planck's measurement of $A_{s}^{\rm obs} = 2.196 \times 10^{-9}$ \cite{Ade:2015xua}. Na\"ively, the operator in Eq.~(\ref{Lint}) ruins the shallow slow-roll potential described in Eq.~(\ref{sixmodels}) by inducing a $\phi^4$ term at  one loop, which is problematically large for $\kappa \gtrsim 4\pi \sqrt \lambda$. For the purposes of this study, however, we only require that $\cL_{\rm int}$ assumes the quadratic form in Eq.~(\ref{Lint}) for small field values, and it is entirely reasonable to assume that it flattens at large values (for example, see Refs.~\cite{Lerner:2009xg,Kallosh:2015lwa}) with a functional form similar to those in Eq.~(\ref{sixmodels}), leading to a negligible contribution during inflation.

Similarly, the dramatic separation of scales in the inflationary potential ($m_\phi \ll \phi_0$) may give the reader pause. However, as the post-inflationary phenomenology is intimately connected to the Higgs sector in our toy model, it is perhaps reasonable to assume that the same as-yet-unknown physics that resolves the Higgs hierarchy problem is also responsible for setting the scales in the inflaton sector. We leave deeper exploration of such model-building questions to future work.




\section{Reheating}

As inflation comes to an end, the universe is filled with an inflaton condensate that transfers its energy to a plasma of Standard Model particles through the process of reheating~\cite{Boyanovsky:1996sv,Bassett:2005xm,Allahverdi:2010xz,Amin:2014eta}. In the class of scenarios considered here, a $Z_2$ symmetry forbids the inflaton from decaying at late times when $\phi = 0$, but this symmetry is spontaneously broken during reheating when $\phi \ne 0$. Consequently, reheating proceeds through a combination of $\phi$ annihilation and $\phi$-dependent decays. Collectively, the non-perturbative description of this evolution is known as preheating~\cite{Kofman:1994rk,Kofman:1997yn}.  

In this section we argue that in the scenarios considered here, reheating efficiently destroys the inflaton condensate, generating a high-temperature plasma of Standard Model particles in thermal equilibrium with inflatons. In the context of the Higgs portal interaction, analytical and numerical studies have shown reheating to be very efficient~\cite{Lerner:2011ge,Bezrukov:2008ut,Ballesteros:2016xej,Enqvist:2016mqj}, easily leading to $T_{\rm RH} \gg m_\phi$.   
As the dynamics of preheating are highly nonlinear and nonperturbative, a detailed study is beyond the scope of this article. In the remainder of this section, we will use a simplified perturbative description of reheating through $\phi$ annihilations to develop the reader's intuition.

Once inflation has ended, the inflaton field begins to oscillate about the minimum of its potential. Since the Higgs mass depends on $\phi$ (see below Eq.~\ref{Lint}), annihilations and decays of $\phi$ are kinematically forbidden for values of $\phi$ for which $V''(\phi) < \kappa \phi^2$, where $V''(\phi)$ and $\kappa \phi^2$ are the field-dependent squared masses of the inflaton and Higgs fields, respectively.  Higgs production becomes efficient for $|\phi| \lsim m_{\phi}/\sqrt{\kappa}$, and a burst of Higgs boson production occurs each time $\phi$ oscillates through the origin of the potential~\cite{Chung:1999ve,Fedderke:2014ura,Enqvist:2016mqj}. The inflaton field reaches the origin on a timescale given by the inverse of the field-dependent inflaton mass, during which Hubble expansion dilutes the energy density by only an order one factor. 

To estimate the efficiency of Higgs production, we treat the inflaton condensate as a collection of zero-momentum particles, and we calculate the cross section for the annihilation channel, $\phi\phi\to\cH^\dagger \cH$.  With the Higgs portal interaction in Eq.~(\ref{Lint}), and for $m_{\phi} \gg m_h$, we obtain the following cross section in the non-relativistic limit:
\begin{equation}
\label{higgs-annihilation}
(\sigma v)_{\phi \phi \to \cH^\dagger \cH} = \frac{\kappa^2}{16 \pi m^2_{\phi}} \, . 
\end{equation}
Inflaton annihilations and decays populate the Standard Model thermal bath, which carries an energy density $\rho_R(t_{\rm RH}) = (\pi^2/30) g_\ast T_{\rm RH}^4$.  We estimate the annihilation rate of inflatons as $\Gamma \sim n_\phi \sigma v$, where $n_\phi$ is the density of the inflaton condensate at the end of inflation and $\sigma v$ is given by Eq.~(\ref{higgs-annihilation}). Because the condensate is very dense, $n_\phi \sim \rho_\phi / m_\phi \gg m_\phi^3$, the annihilation rate dramatically exceeds the Hubble expansion rate, $\Gamma \gg H$, once the annihilation and decay channels are no longer blocked.  This na\"ive, perturbative estimate suggests that reheating is very efficient, leading to a rapid transfer of energy from the inflaton condensate to the Standard Model thermal bath.  A more accurate description of reheating would take into account other relevant factors~\cite{Kofman:1997yn}. In particular, annihilations occur over a short time interval, for which the uncertainty principle bounds $E_\phi \sim 1 / \Delta t \gg m_\phi$, broadening the range of field values for which annihilations can occur while also suppressing the annihilation cross section. On the other hand, the inflaton condensate is very dense, enabling many-to-two annihilation processes to proceed efficiently. 
 
We can identify the reheating temperature as follows:
\beq
T_{\rm RH} = \left(\xi  \frac{30}{\pi^2 g_*} V(\phi_{\rm end}) \right)^{1/4} \, , 
\eeq
where $V(\phi_{\rm end})$ is the energy density at the end of inflation, and we have included the quantity $\xi$ to account for the possibility that energy transfer is not perfectly efficient during reheating. From our na\"ive perturbative calculation, we expect $\xi =\mathcal{O}(1)$ as found, for instance, in instant preheating scenarios~\cite{Felder:1998vq}. Numerical studies of reheating in other models obtain $\xi \sim \mathcal{O}(0.001-0.1)$ \cite{Kofman:1997yn,Felder:1998vq,Dufaux:2007pt,Hyde:2015gwa}. In the remainder of this article we will assume that the universe reheats to a high temperature, $T_{\rm RH} \gg m_{\phi}$, and that the inflaton thermalizes with the Standard Model plasma. 
 



\section{Dark Matter Phenomenology}
\label{DM}

\begin{figure}[t!]
\hspace{-0.in}
\includegraphics[width=3.3in,angle=0]{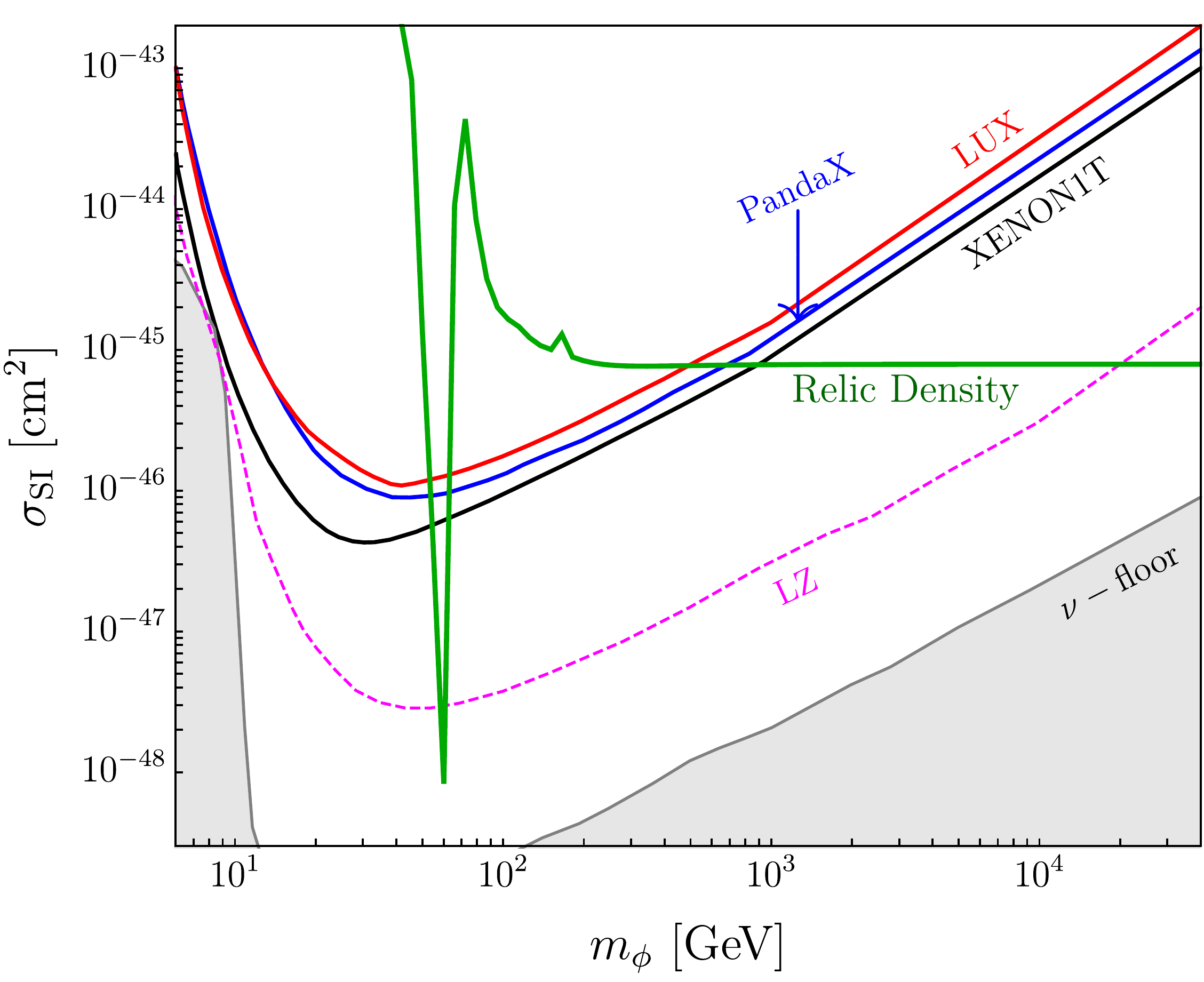}
\caption{The spin-independent elastic scattering cross section with nucleons of inflaton dark matter, compared to the constraints from the XENON1T~\cite{Aprile:2018dbl}, LUX~\cite{Akerib:2016vxi} and PandaX-II~\cite{Cui:2017nnn} Collaborations. We also show the projected reach of the LZ Collaboration~\cite{Akerib:2015cja,Cushman:2013zza}, as well as the neutrino floor~\cite{Billard:2013qya}.}
\label{direct-detection}
\end{figure}

After reheating, the inflaton abundance follows the equilibrium number density until thermal freeze-out occurs, at a temperature given by $T_{\rm FO} \simeq m_\phi/20$~\cite{Kolb:1990vq}. Using the expression for the annihilation cross section given in Eq.~(\ref{higgs-annihilation}), supplemented with correct kinematic factors and including all annihilations channels, we find that the inflaton relic abundance is
\begin{eqnarray}\label{relic}
\Omega_{\phi} h^2 \approx 0.1 \times \bigg(\frac{0.3}{\kappa}\bigg)^2 \, \bigg(\frac{m_{\phi}}{{\rm TeV}}\bigg)^2 \, . 
\end{eqnarray}
Thus the measured dark matter abundance ($\Omega_{\rm DM}h^2 \simeq 0.12$~\cite{Ade:2015xua}) can be obtained for reasonable choices of parameters. This interaction also leads to the following spin-independent scattering cross section per nucleon:
\be
\label{sigma-dd}
\sigma_{\rm SI} \simeq     \frac{ \kappa^2  v^2 m^2_n}{16\pi m_h^4 m^2_{\phi}}  \, \bigg[ \frac{Z}{A} C_{p} + \bigg(1-\frac{Z}{A}\bigg) C_{n} \bigg]^2,
\ee
where $Z$ and $A$ are the atomic number and atomic mass of the target nucleus, $v=246$ GeV, and $C_{p,n}$ are the effective couplings to nucleons~\cite{Shifman:1978zn,DelNobile:2013sia}. For $m_{\phi} \gg m_h$ and the case of a xenon target
\be \label{sigma-dd-val}
\sigma_{\rm SI} \approx 7 \times 10^{-46} \, {\rm cm}^2 \, \times \bigg(\frac{\kappa}{0.3}\bigg)^2 \, \bigg(\frac{{\rm TeV}}{m_{\phi}}\bigg)^2.
\ee
In Fig.~\ref{direct-detection}, we compare this cross section to the constraints placed by the XENON1T~\cite{Aprile:2018dbl}, LUX~\cite{Akerib:2016vxi} and PandaX-II~\cite{Cui:2017nnn} Collaborations. We also show the projected reach of the LZ Experiment~\cite{Akerib:2015cja,Cushman:2013zza}, as well as the neutrino floor below which such experiments encounter an irreducible background from neutrino interactions~\cite{Billard:2013qya}. Much like other Higgs-portal dark matter scenarios~\cite{Escudero:2016gzx,Silveira:1985rk,Patt:2006fw,McDonald:1993ex,Burgess:2000yq,Pospelov:2007mp,MarchRussell:2008yu,Gonderinger:2012rd}, the dark matter in this model is constrained by this class of experiments to approximately $m_{\phi} \gsim 900$ GeV (except near $m_{\phi} \approx m_h/2)$.

 
Although we have focused here on inflaton dark matter with a Higgs portal coupling, we could have considered other interactions to facilitate dark matter annihilation and local dark matter detection. For example, a simple variation could involve inflatons with a mixed quartic coupling with a scalar that is a Standard Model singlet, allowing the dark matter to annihilate to pairs of ``dark Higgses,'' which then decay to Standard Model particles. In this variation, direct detection constraints could be significantly relaxed, and thereby allow for lighter inflatons and improved prospects for indirect searches~\cite{Escudero:2017yia,Liu:2014cma,Ko:2014loa,Abdullah:2014lla,Martin:2014sxa,Berlin:2014pya,Hooper:2012cw}. One could also consider models in which the inflaton annihilates to a pair of dark photons or right-handed neutrinos~\cite{Alexander:2016aln}.




\section{Summary and Conclusions}

In this letter, we have considered a range of WIMPflation scenarios, in which the inflaton also serves as a viable dark matter candidate. Requiring the inflaton to be stable implies that reheating must be accomplished through a combination of annihilations and background-field-dependent decays. The same annihilation process later sets the thermal relic abundance of inflaton dark matter. We identify several inflationary potentials that can accommodate all current CMB constraints and yield a viable thermal relic abundance of inflaton dark matter. For the case of inflaton dark matter that annihilates through a Higgs portal coupling, we find encouraging prospects for future direct detection experiments.

\begin{acknowledgments}  
We would like to thank Mark Hertzberg for helpful discussions. AJL is supported by the University of Chicago by the Kavli Institute for Cosmological Physics through grant NSF PHY-1125897 and an endowment from the Kavli Foundation and its founder Fred Kavli. This manuscript has been authored by Fermi Research Alliance, LLC under Contract No. DE-AC02-07CH11359 with the U.S. Department of Energy, Office of Science, Office of High Energy Physics. The United States Government retains and the publisher, by accepting the article for publication, acknowledges that the United States Government retains a non-exclusive, paid-up, irrevocable, world-wide license to publish or reproduce the published form of this manuscript, or allow others to do so, for United States Government purposes.
\end{acknowledgments}

\bibliography{WIMPflation}

\end{document}